\def\bold#1{\setbox0=\hbox{$#1$}%
     \kern-.025em\copy0\kern-\wd0
     \kern.05em\copy0\kern-\wd0
     \kern-.025em\raise.0433em\box0 }
\def\slash#1{\setbox0=\hbox{$#1$}#1\hskip-\wd0\dimen0=5pt\advance
       \dimen0 by-\ht0\advance\dimen0 by\dp0\lower0.5\dimen0\hbox
         to\wd0{\hss\sl/\/\hss}}
\documentstyle[12pt]{article}

\newlength{\dinwidth}
\newlength{\dinmargin}
\newcommand{\resection}[1]{\setcounter{equation}{0}\section{#1}}

\begin{document}

\def\lq{\left [}
\def\rq{\right ]}
\def\LL{{\cal L}}
\def\VV{{\cal V}}
\def\AA{{\cal A}}
\def\MM{{\cal M}}
\def\eps{\epsilon}

\newcommand{\be}{\begin{equation}}
\newcommand{\ee}{\end{equation}}
\newcommand{\bea}{\begin{eqnarray}}
\newcommand{\eea}{\end{eqnarray}}
\newcommand{\nn}{\nonumber}
\newcommand{\dd}{\displaystyle}

\thispagestyle{empty}
\vspace*{4cm}
\begin{center}
  \begin{Large}
  \begin{bf}
THEORY REVIEW OF CP VIOLATIONS IN B DECAYS AT HADRON COLLIDERS
\\
  \end{bf}
  \end{Large}
  \vspace{15mm}
  \begin{large}
G. Nardulli\\
  \end{large}
Dipartimento di Fisica, Universit\'a di Bari\\
I.N.F.N., Sezione di Bari, Italy\\
\end{center}
  \vspace{15mm}
\begin{center}
BARI-TH/93-136\\
\vspace*{10mm}
{\it Talk given at the UNK B factory workshop\\
Liblice Castle, Czechoslovakia, January 1993}
\end{center}
\vspace*{15mm}
\begin{center}
  \begin{Large}
  \begin{bf}
  ABSTRACT
  \end{bf}
  \end{Large}
\thispagestyle{empty}
\begin{quotation}
  \vspace{5mm}
\noindent
CP violations in B meson decays at hadron colliders are reviewed. In particular
I examine: $B^0-\bar{B^0}$ mixing and oscillations within Standard Model and
the Cabibbo-Kobayashi-Maskawa picture for CP violations; $B^0$ decays
into CP eigenstates and the measurements of the angles in
the unitarity triangle; finally I consider a class of
charged $B$ decays that allow
to avoid uncertainties arising from the computation of
hadronic matrix elements.
\end{quotation}
\end{center}
\newpage
\setcounter{page}{1}
\resection{Introduction}
The discovery of $CP$ violations in the kaon system \cite {cronin} was
no doubt a remarkable breakthrough in elementary particle physics;
nonetheless, despite its fundamental importance, $CP$ violations appear to
be a phenomenon not well understood from a theoretical point of view
and not well documented experimentally, since, after almost 30
years from its discovery, it has been observed only in kaon systems.

B physics offers the possibility to study $CP$ violations in
a different context and therefore promises to provide a clue for
a better understanding of this phenomenon. Indeed it would
allow to test the
predictions of the Standard Model, based on the Cabibbo - Kobayashi -
Maskawa (CKM) picture of the weak mixing \cite {ckm}, and maybe to discover new
physics beyond the Standard Model.

High energy high luminosity hadron colliders, such as
the future Large Hadron Collider (LHC), the Super Conducting
Super Collider (SSC) or UNK, are particularly suitable
to perform these investigations, because of
the copious production of  B's that is expected. In this paper I review the
most
promising channels for observing CP violations in $B$ decays and for testing
the predictions of the Standard Model (for other reviews on this subject
see \cite {reviews}). The paper is organized as follows. In Section 2
I review $B^0 - \bar {B^0}$ mixing and oscillations while  in Section 3 the
CKM picture of weak mixing within the Standard Model is examined.
Section 4 is devoted to the analysis of $CP$ violations for
neutral $B$ decays into $CP$ eigenstates and Section 5 contains a
discussion of some decay channels of charged B's. Finally Section 6
contains the conclusions.
\resection{$B^0 - \bar {B^0}$ mixing and oscillations}

As well known, the $B^0$ (=
$d{\bar b}$) and $\bar {B^0}$ (=$b{\bar d}$) mesons
mix with each other due to 2nd order flavour violating weak interactions.
This phenomenon is similar to the $K^0 - \bar {K^0}$ mixing and
has been observed experimentally by both the CLEO \cite {cleo}
and ARGUS \cite {argus}
 Collaborations in $e^+ - e^-$ scattering
at the energy of $\Upsilon (4S)$ .
The present experimental situation can be summarized as follows. If
one defines
\be
r=\frac{N(B^0 B^0) - N(\bar {B^0} \bar {B^0})}{N(B^0 \bar {B^0})
+ N(\bar {B^0} B^0)}
\ee
where, e.g., $N(\bar {B^0} \bar {B^0})$ is the number of
$\bar {B^0} \bar {B^0}$ pairs observed in $e^+ e^-$  collisions at
$\Upsilon(4S)$, then one has:
\be
x_d={\sqrt \frac{2 r}{1-r}}=0.66 \pm 0.11~.
\ee
Since $b$ and $\bar b$ are always produced in pairs in
$e^+ e^-$  collisions , $r \not= 0$ (or $x_d \not= 0$) means that the $b$
quark (or $\bar b$ antiquark) has changed to $\bar b$ (resp. $b$). This is
a second order  weak interactions process since it violates
beauty by two units ($\Delta B=2$).

The two-state $B^0 \bar {B^0}$ system, as other similar physical systems,
can be generally described by a $2 \times 2$ matrix Hamiltonian \cite {wigner}
\be
H~={\bf M} - \frac{i {\bf \Gamma}}{2} =
\left (\begin{array}{cc}
M -i \Gamma/2 & M_{12} -i \Gamma_{12}/2 \nn\\
M^*_{12} -i \Gamma^*_{12}/2 &M -i \Gamma/2
\end{array}\right )
\ee
where ${\bf M}$ and ${\bf \Gamma}$ are hermitean matrices and we work
in the basis where ${\displaystyle B^0 =\left (\begin{array}{c}1\\0
\end{array}\right )}$
and ${\displaystyle \bar {B^0}=\left (\begin{array}{c}0\\1
\end{array}\right )}$. In (2.3) we have put
${\bf M_{11}}={\bf M_{22}}=M$  and ${\bf \Gamma_{11}}= {\bf \Gamma_{22}}
=\Gamma$ because of $CPT$ invariance. We stress that $H$ contains
non vanishing off-diagonal matrix elements due to weak interactions.

We now call $B_1$ and $B_2$ the eigenstates of $H$. They can
be written as superpositions of the $B^0$ and $\bar {B^0}$ states;
therefore one has
\be
B_{1,2}=\frac{(1+\eps)B^0 \pm (1-\eps)\bar {B^0}}{\sqrt {2(1+|\eps|^2)}}
\ee
corresponding to the eigenvalues of $H$:
\be
H_1~= M_1 -i \frac{\Gamma_1}{2} = M + Re Q - \frac{i}{2}(\Gamma - 2 Im Q)
\ee

\be
H_2 = M_2 -i \frac{\Gamma_2}{2} = M - Re Q - \frac{i}{2}(\Gamma + 2 Im Q)
\ee
with
\be
\frac{1 -\eps}{1 + \eps} = \hat \eta = {\sqrt \frac {M^*_{12} -
i \Gamma^*_{12}/2}{M_{12} -i \Gamma_{12}/2}}
\ee
and
\be
Q = (M_{12} -i \Gamma_{12}/2) {\hat \eta} = \sqrt{ (M_{12} -i \Gamma_{12}/2)
(M^*_{12} -i \Gamma^*_{12}/2)}
\ee

Since $H_1 \not= H_2$, the time evolutions of the mass
eigestates
$B_1$ and $B_2$ (that are given by $\sim e^{-i H_j
t} |B_j(0)>$) are different. Therefore one can have oscillations
between $B^0$ and $\bar {B^0}$, analogous to $K^0 - \bar {K^0}$
oscillations. As a matter of fact, denoting by $|B^0(t)>$
 ($|\bar {B^0}(t)>$) the state
that at the time $t=0$ is pure
$B^0$ (resp $\bar {B^0}$), one has
\bea
|B^0(t)> &=& \frac{1}{2}[e^{- i M_{1}t - \Gamma_{1}t/2}
 + e^{- i M_{2}t - \Gamma_{2}t/2}]|B^0>+\nn\\
 &+& \frac{\hat \eta}{2}[e^{- i M_{1}t - \Gamma_{1}t/2}
 - e^{- i M_{2}t - \Gamma_{2}t/2}]|\bar {B^0}>
\eea
and
\bea
|\bar {B^0}(t)> &=& \frac{1}{2 \hat \eta}[e^{- i M_{1}t - \Gamma_{1}t/2}
 - e^{- i M_{2}t - \Gamma_{2}t/2}]|B^0>+\nn\\
&+& \frac{1}{2}[e^{- i M_{1}t - \Gamma_{1}t/2}
 + e^{- i M_{2}t - \Gamma_{2}t/2}]|\bar {B^0}>
\eea

One can extract information on the parameters $M_j,~~ \Gamma_j$
appearing in previous equations, by a theoretical analysis of the
box diagram \cite {gaillard} \cite {altarelli} which
is a weak interaction graph where two $W$'s are excanged between the
quarks (see Fig 1). Indeed, from (2.3),
$M_{12}$ coincides with the real part of $H_{12}$ ($M_{12}=Re H_{12}=
Re~<B^0| H |\bar {B^0}>$)
and $-\Gamma/2$ with its imaginary part (neglecting $CP$
violations). In computing $Re~ H_{12}$, the largest contribution arises
when the quarks on the internal lines of the box diagrams are top quarks;
therefore $M_{12}$ will be approximately given by
\be
M_{12}\sim (V_{tb}V^*_{td})^2m_t^2
\ee
since the charm quark contribution is depressed by a factor $(m_c/m_t)^2$.
On the other hand, in the computation of
$\Gamma_{12}$ by Landau-Cutkosky rules,
which corresponds
to replace in the propagators $(p^2-m^2)^{-1}$ by
$- 2 \pi i \delta_+ (p^2-m^2 )$, one has to cut internal lines,
and therefore to consider decays into real particles.
In these decays the energy of the final state cannot excede $m_B$
(in the $B$ meson rest frame). Therefore in this case the internal quarks in
the box diagrams can only be up or charm and one has
\be
\Gamma_{12} \sim m^2_b(V_{cb}V^*_{cd} + V_{ub}V^*_{ud})^2 =
(V_{tb}V^*_{td})^2 m^2_b
\ee
where the last equality follows from the unitarity of the CKM matrix and
we have neglected light quark masses as compared to $m_b$; we stress that
no $m_t^2$ factor can be generated by the computation of $\Gamma_{12}$.

{}From previous equations we have the following consequences. First,
from (2.11) and (2.12):
\be
|M_{12}|>>|\Gamma_{12}|
\ee
since $m_t>>m_b$; moreover $M_{12}$ and $\Gamma_{12}$ have the same phase:
\be
\Phi=arg ~(M_{12})=arg~( \Gamma_{12})~ .
\ee
Furthermore, from (2.7), (2.13) and (2.14) it follows that
\be
Q= \sqrt{ (M_{12} -i \Gamma_{12}/2)
(M^*_{12} -i \Gamma^*_{12}/2)}
= |M_{12}|- i \frac{|\Gamma_{12}|}{2}\approx |M_{12}|;
\ee
therefore $Im~ Q$ is negligible and, from (2.5), one has
\be
\Delta \Gamma \equiv \Gamma_1 - \Gamma_2 \simeq 0~.
\ee
In other words the states $B_1$ and $B_2$ have the same lifetime (this is at
at variance with the $K^0 - \bar {K^0}$ mixing where
the states $K_L,~ K_S$ have very different decay widths).

Finally, from (2.7), (2.13) and (2.14) one gets:
\be
\hat \eta = \frac{1- \eps}{1+ \eps} \simeq e^{-i \Phi} =
\frac{(V^*_{tb}V_{td})^2}{|V^*_{tb}V_{td}|^2}~.
\ee

After having fixed the phase factor $\hat \eta$, we now turn to
the mixing parameter $x_d$ in (2.2). Its definition is as follows:
\be
x_d=\frac {\Delta m}{\Gamma},
\ee
with
\be
\Gamma \equiv \frac{\Gamma_1 + \Gamma_2}{2} = \frac{1}{\tau_{B^0}}
\ee
and
\bea
\Delta m &=& M_1 - M_2 = Re (H_1 - H_2) = 2 Re Q\nn\\
&=& 2 Re <B^0|H_{eff}|\bar {B^0}>~.
\eea

The value of the parameter $x_d$ (or equivalently $\Delta m$) can be
obtained by computing the box diagram; this is done
by a method analogous to
that originally employed by Gaillard and Lee \cite {gaillard} for
$K^0 - {\bar K^0}$ mixing. In the present case the box diagram produces
the effective hamiltonian
\be
H_{eff}=\frac{G^2}{16 \pi^2} m^2_t (V^*_{td}V_{tb})^2 A(x)
:\bar d \gamma_\mu(1-\gamma_5)b \bar d \gamma^\mu(1-\gamma_5)b:
\ee
where $x=m_t^2/m_W^2$ and $A(x)$ is a smooth decreasing function
assuming the values $A \approx 0.75$ (for $m_t = m_W$) and $A \sim 0.5$
(for $m_t \sim 250 GeV$).

{}From (2.20) and (2.21) one gets the formula:
\be
x_d=\frac{G^2}{6 \pi^2} \tau_{B^0}   m^2_t |V^*_{td}V_{tb}|^2 A(x)
B_B f_B^2,
\ee
where $f_B$ is the leptonic $B$ decay constant, defined by
\be
<0|\bar d \gamma_\mu \gamma_5 b|\bar {B^0}(p)> = i p_\mu f_B
\ee
and $B_B$ is defined through the relation
\be
<B^0(p)|\bar d \gamma_\mu(1-\gamma_5)b \bar d \gamma^\mu(1-\gamma_5)b
|\bar {B^0}(p)> = \frac{8}{3} B_B m^2_B f^2_B \frac{1}{2 m_B},
\ee
which means that $B_B \not= 1$ would signal deviations from the factorization
approximation.

We shall discuss the numerical values for $f_B$ and $B_B$ in the next Section.

\resection{The CKM picture}

A simple example will show a typical mechanism of $CP$ violation within the
Standard Model.

Suppose we want to measure $CP$ violation in the charged B decay
\be
B^+ \to K^+ \rho^0
\ee
Quark diagrams with different topologies have been classified
in \cite {linglichau}. In the present case
they produce the amplitude:
\be
A (B^+ \to K^+ \rho^0) = V^*_{ub} V_{us} A_1 + V^*_{cb} V_{cs} A_2.
\ee
Among the quark diagrams contributing to $A_1$  we have the u-spectator
diagram (also called external W-emission;
we adopt here the terminology of \cite {linglichau}), whereas $A_2$ arises
from the penguin charm-diagram.
Let us now consider the charged conjugate process:
\be
B^- \to K^- \rho^0
\ee
whose amplitude is given by
\be
A (B^- \to K^- \rho^0) = V_{ub} V^*_{us} A_1 + V_{cb} V^*_{cs} A_2,
\ee
where the strong amplitudes $A_j$ have not changed since strong interactions
conserve $CP$.
The $CP$ violating asymmetry for this process is
\bea
\Delta &=& \frac{\Gamma(B^+ \to K^+ \rho^0)  - \Gamma(B^- \to K^- \rho^0)}
{\Gamma(B^+ \to K^+ \rho^0)  + \Gamma(B^- \to K^- \rho^0)} \nn\\
&=& \frac {- 4 Im [V^*_{ub} V_{us}  V_{cb} V^*_{cs} ]\times Im A_1 A^*_2}
{|V^*_{ub} V_{us} A_1 + V^*_{cb} V_{cs} A_2|^2 +
|V_{ub} V^*_{us} A_1 + V_{cb} V^*_{cs} A_2|^2}
\eea
which shows that in order to have $CP$ violations one needs two phases:
one is provided by strong interactions (indeed we
need two strong amplitudes with $Im A_1 A^*_2 \not= 0$)
and the other one from weak
interactions ($Im [V^*_{ub} V_{us}  V_{cb} V^*_{cs} ] \not= 0$).

The Standard Model with three families of quarks produces quite naturally
the
weak phase. As well known, the CKM matrix stems from the
fact that the weak eigenstates, i.e. states having definite transformation
properties under the gauge group $SU(2)\times U(1)$, in general
are not mass
eigenstates. The matrix
that relates weak and mass eigenstates is indeed the CKM matrix,
whose entries $V_{\alpha j}$, multiplied by $G/\sqrt 2$, give
the coupling of the weak charged current $\bar q_{\alpha} \gamma_\mu (1-
\gamma_5)q_j$.
The CKM matrix can be written as follows:
\bea
{V}=
\left (\begin{array}{ccc}
V_{ud} & V_{us} & V_{ub}\\
V_{cd} & V_{cs} & V_{cb}\\
V_{td} & V_{ts} & V_{tb}
\end{array}\right )
\eea
and is unitary:
\be
V^\dagger V= V V^\dagger =1
\ee

It is known that a $3\times 3$ unitary matrix,
after removal of the unphysical quark field phases,
 depends on 4 parameters, while an
orthogonal  $3\times 3$ matrix, i.e. a unitary  $3\times 3$
matrix with real coefficients,
depends on 3 parameters. Since we do not expect
that $V_{\alpha j}$ is orthogonal, in general $V_{\alpha j}$
should be
truly complex, i.e. it should contain a phase.

A useful way to represent $V_{\alpha j}$ is given by
the Wolfenstein parametrization \cite {wolfenstein}
\bea
{V}=
\left (\begin{array}{ccc}
1- \lambda^2/2 & \lambda & A \lambda^3 [\rho - i \eta (1 - \lambda^2/2)]\\
-\lambda & 1- \lambda^2/2 - i \eta A^2 \lambda^4 & A \lambda^2
(1 + i \eta \lambda^2)\\
A \lambda^3 (1 - \rho - i \eta) & - A \lambda^2 & 1
\end{array}\right )
\eea

$\lambda$ is given by $sin \theta_c$ ($\theta_c$ the Cabibbo angle):
$\lambda = sin \theta_c = 0.221$, whereas $A$ and $\rho$ are
numbers of the order 1 to be discussed below and $\eta$
is related to the weak $CP$ violating phase.

Before discussing the experimental constraints on the parameters
in (3.8), let us discuss a few properties of the CKM matrix.

An immediate consequence of unitarity is the formula \cite {jarlskog}
\be
Im [V_{\alpha j}~V_{\beta k}~V^*_{\alpha k}~V^*_{\beta j}]~ =~ J~
\sum_{\gamma, l} \eps_{\alpha \beta  \gamma} \eps_{jkl}
\ee
where $J$ is independent of the parametrization used for
the CKM matrix and of phase conventions and in
the Wolfenstein parametrization is given by:
\be
J=\eta A^2 \lambda^6
\ee

Another important feature arising from the unitarity
of V is the geometrical interpretation of the relations:
\be
(V^\dagger V)_{n m} =\sum_{\gamma } V^*_{\gamma n} V_{\gamma m} =~0~~~
{}~~(n \not= m)
\ee
as the zero sum of three vectors in the complex plane which, therefore,
set the the
border of a triangle (unitarity triangle).

We shall consider here three of these relations:
\be
V_{u d}V^*_{u b}~+~V_{c d}V^*_{c b}+~V_{t d}V^*_{t b}~=~0
\ee
\be
V_{u d}V^*_{u s}~+~V_{c d}V^*_{c s}+~V_{t d}V^*_{t s}~=~0
\ee
\be
V_{u s}V^*_{u b}~+~V_{c s}V^*_{c b}+~V_{t s}V^*_{t b}~=~0
\ee
One can easily show that the parameter $J$ in (3.9) has the
geometrical interpretation of $2 \times$ (area of the unitarity triangle);
therefore all the triangles defined by (3.12)-(3.15) have the same area.
Incidentally, from (3.5) we see that also the $CP$ violating asymmetry
$\Delta$ in $B^{\pm} \to K^{\pm} \rho^0$ is proportional to the
area of the unitarity triangle.

Whereas all the unitarity triangles have the same area, their shape can
differ significantly. Indeed from (3.8) we see that in the triangle
(3.12) all the sides are of the same order ($\sim \lambda^3$);
on the contrary, in the
triangle (3.13) one side is very small ($|V_{t d}V^*_{t s}| \sim \lambda^5$)
and the other ones are of the order of $\lambda$, while in the triangle (3.14)
one side is small, with length $\sim \lambda^4$ and the other two are
$\sim \lambda^2$. The different shape is related to different physical
properties: in (3.12) the smallness of all the sides reflects the relatively
large $B$ lifetime; in (3.13) the smallness of one side reflects
the low value of the $CP$ violating parameter $\eps_K$ in $K-$decay,
while the small side in (3.14) results in the small asymmetries in the decays
in $B_s \to \psi \phi, ~ \eta_c \phi$ (see below).

Let us now discuss in more detail (3.12); it is worth stressing that
in the literature the name "unitarity triangle" is
generally reserved only to the triangle associated to this relation.
We put $V_{c d}V^*_{c b}$ on
the real axis in the  $(\rho , \eta)$
complex plane and scale down all the sides of the
unitarity triangle by $|V_{c d}V^*_{c b}|$. In the Wolfenstein
parametrization the three vertices of the scaled triangle have coordinates:
\be
A~=~(\rho , \eta)~~~~~~~~~~~~~B~=~(1,~0)~~~~~~~~~~~~~C~=~(0,~0)
\ee
and the corrispondent angles will be denoted by $\alpha, \beta, \gamma$
respectively (see Fig. 2).
It is clear that $CP$ violations ($\eta \not= 0$) in  $B$ physics within
the Standard Model are only possible if all the three angles are different
from $0$. Indeed it is easy to prove the relations:
\be
tg~\beta ~=~\frac{\eta}{1-\rho}~~~~~~~~~tg~\gamma ~=~\frac{\eta}{\rho}
\ee
and
\bea
sin ~ 2\beta &=&\frac{2 \eta(1 -\rho)}{(1-\rho)^2+\eta^2}\nn\\
sin ~ 2\gamma &=&\frac{2 \eta \rho}{\rho^2+ \eta^2}
\eea

Let us conclude this Section by discussing experimental constraints
on the parameters of the CKM matrix.
As stated above, from semileptonic light hadron decays we have \cite {pdb}
\be
\lambda~=~0.221
\ee
A is related to $|V_{cb}|$ by
the formula $|V_{cb}| \simeq A \lambda^2$, and $|V_{cb}|$
can be extracted by the semileptonic $B$ decays, using theoretical
information on the form factors coming from QCD sum rules \cite {pavernoi}
or the Heavy Quark Effective Theory \cite {neubert}
with an appropriate
Isgur-Wise \cite {IW} universal function (for reviews see
\cite {nardulli}). We quote the "best value" of this
analysis
\be
|V_{cb}| = 0.044 \pm 0.009
\ee
which leads to
\be
A=0.90\pm 0.10
\ee
Next we consider constraints from
$|V_{ub}/V_{cb}|$. The value of
$|V_{ub}|$ can be obtained by the analysis of the lepton energy
spectrum at the end-point as suggested by \cite {acmor}.
{}From this analysis the CLEO \cite {cleo2} and
ARGUS \cite {argus2} Collaborations get a signal that is
interpreted as $|V_{ub}| \not= 0$. We use the result \cite {cleo2}
\be
|V_{ub}/V_{cb}| = 0.09 \pm 0.04~,
\ee
which, together with the relation
\be
|V_{ub}/V_{cb}| \sim \lambda {\sqrt{ \rho^2 + \eta^2}} ~,
\ee
defines in the $\rho - \eta$ plane a region between two half-circles
centered at $(0,0)$ and having different radii.

Another piece of information comes from the $ B^0 - \bar {B^0}$ mixing.
Eq. (2.23) relates $x_d$ to $|V_{tb}V^*_{td}|^2$. Since
$|V_{tb}V^*_{td}| \simeq
A \lambda^3 {\sqrt { (1 - \rho)^2 + \eta^2}}$, the
experimental result (2.2) defines a region in the
$\rho - \eta$ plane included between two half-circles centered at
$(\rho , \eta)~=~(1,0)$. This region depends also on the value of
the top quark mass and on the parameters appearing in (2.22), most
notably the $B$ meson leptonic decay constant $f_B$
defined in (2.23) and the $B_B$ parameter of (2.24) and we
shall now briefly discuss the value of these two
hadronic parameters.

There has been a large amount of theoretical activity on $f_B$ in the
last few years and
different theoretical approaches
have been used, e.g. QCD sum rules \cite {pavernoi}, relativistic
potential models \cite {pietroni}, lattice QCD \cite {lattice} (for a review
see \cite {nardulli}). We shall adopt here
a rather conservative range of values, already employed by
fromthe LCH B-physics working group \cite {lhc}, i.e.
\be
f_B=220\pm 52 MeV~~.
\ee
The bigger half-circle in Fig. 3 corresponds to the lower value of $f_B$.

As to $B_B$, an analysis of this parameter based on a dispersive
calculation \cite {colpaver} gives
\be
B_B\simeq 1
\ee
which
is also the value commonly used in the literature. Finally
(2.22) depends on the B lifetime, for which \cite {lhc} use the results
of the analysis \cite {gilmannir}, i.e. $\tau_b |V_{cb}|^2=(3.5\pm0.6)~ 10^9
GeV^{-1}$.

The last experimental input
that can be used to
constrain the parameters $\rho$ and $\eta$ comes
from $CP$ violation in $K$ decay (the $\eps_K$ parameter,
which is experimentally given by
$\eps_K=2.27\times 10^{-3}$). The relevant
expression for $\eps_K$ in the Standard Model is \cite {gilmanwise}
\bea
\eps_K &=& \frac{B_K G^2 f_K^2 m_K m^2_c}{6 {\sqrt 2} \pi^2 [m_L-m_S]}
 A^2 \lambda^7 \eta\nn\\
&\times &[-\eta_1 + \eta_3 ln\frac{m_t^2}{m_c^2} + \eta_2
f(\frac{m_t^2}{m_W^2})
\frac{m_t^2}{m_c^2}(A^2\lambda^4 + A^2 \lambda^6 \rho)] e^{i\pi /4}
\eea
where $\eta_j$ are QCD coefficients, $m_L, m_S$ are the $K_L$ and
$K_S$ masses and the function $f$ is  given in \cite {gilmanwise}.
A major source of uncertainty in the previous expression is in the
factor $B_K$ defined analogously to (2.24), i.e.
\be
<K^0(p)|\bar d \gamma_\mu(1-\gamma_5)s \bar d \gamma^\mu(1-\gamma_5)s
|{\bar K^0}(p)> = \frac{8}{3} B_K m^2_K f^2_K \frac{1}{2 m_K}
\ee
$B_K$ has been object of several theoretical investigations,
by chiral symmetry \cite {don}, QCD sum rules (see \cite  {colnar} and
references therein), dispersion relations \cite {cea} and
Lattice QCD \cite  {sharpe}. We now expect a value
of $B_K$ consistent with vacuum saturation ($B_K \approx 0.8 -1.0$),
but again we assume a conservative viewpoint
and, following the LHC study group on B physics
\cite {lhc}, we take
\be
0.33<B_K<1.
\ee

{}From (3.25) and (3.27) and from the experimental value of $\eps_K$ we obtain
a
region included between two
hyperbolae in the $\rho ~ \eta$ plane (the upper
curve corresponds to the lowest value of $B_K$).

The intersection among the allowed regions defined above is depicted in
Fig. 3 for $m_t= 140~ GeV$. The size of the region depends on $m_t$
and a more complete analysis, considering
different value of $m_t$ can be found in \cite {lhc}.

\resection{$B^0$ decays into $CP$ eigenstates final states }
$B^0$ decays into $CP$ eigenstates offer the possibility to observe $CP$
violations     and measure the phase of the CKM matrix without the large
uncertainties typical of the calculations of hadronic quantities.
We shall therefore in this Section focus on
the most promising decay channels into $CP$ eigenstates.

To begin with, we shall consider the measurement of the angle $\beta$
in the unitarity triangle.

{\it Measurement of $sin 2\beta$}

Let us consider the decay
\be
B^0 \to \psi K_S
\ee
whose final state is $CP$ eigenstate with eigenvalue $CP=-1$. The
predicted branching ratio for this decay is
\be
BR(B^0 \to \psi K_S) \simeq 3.6 \times 10^{-4}
\ee
since \cite {pdb}
\bea
BR(B^0 \to \psi K^0) &=& 6.5 \pm 3.1 \times 10^{-4}\\
BR(B^+ \to \psi K^+) &=& 7.7 \pm 2.0 \times 10^{-4}
\eea

The results (4.3)and (4.4)
can be easily accounted for by an internal W-emission
 diagram, which is proportional to the product $V^*_{cb} V_{cs}$.
Indeed, by computing it in the factorization approximation, one gets (with
$|V^*_{cb}|= 0.044$) \cite {radiative}
\be
BR(B^+ \to \psi K^+) = 9.1 \times 10^{-4}
\ee
which agrees with (4.3), (4.4)
within the experimental errors. Therefore according
to this analysis  \cite {radiative} $BR(B^0 \to \psi K_S) \simeq 4.5
\times 10^{-4}$.

The internal W emission diagram has zero weak phase since $V^*_{cb} V_{cs}$
is almost real (see (3.8)).
However, according to the discussion in Section 2, the
state $|B^0>$ at time $t=0$ will contain, at time $t \not= 0 $, an
admixture of both $|B^0>$ and $|\bar {B^0}>$, with a mixing parameter having
  a weak phase  through $\hat \eta$ (see (2.17)). Therefore,
by this mechanism, we have again an interference between two
amplitudes, with different weak phases, one arising from direct
$B^0 \to \psi K_S$ decay and the other one arising
because of the mixing $B^0 - \bar {B^0}$, i.e. by
the process $B^0 \to \bar {B^0} \to \psi K_S$.

For the amplitudes describing these decays we can write:
\bea
<\psi(p',\eps ) K_S(q) | L | B^0(p) > &=&~ a (\eps^* \cdot p)\\
<\psi(p',\eps ) K_S(q) | L | \bar {B^0}(p) > &=&~ a (\eps^* \cdot p)
\eea
Indeed we observe that, in general, if $|f>$ is $CP$ eigenstate with
$CP=\pm 1$, then $<f| L | B^0>~ =~ \pm e^{-i \alpha}
<f| L' | \bar {B^0}>$,  where $ CP~ | B^0>~ =~e^{-i \alpha}
| \bar {B^0}>$  ($\alpha = $ arbitray phase) and $ L'$ is
the $CP$ transformed lagrangian. In (4.6) and (4.7) we have
 put  $L =L'$ because strong interactions
preserve $CP$ and $V^*_{cb} V_{cs}$ is real
and we have choosen the phase convention
$ CP~ | B^0>~ =~ -
| \bar {B^0}>$, i.e. $\alpha = \pi $.

If we now compute (4.6)-(4.7) at time $t$, from (2.9)-(2.10) we get
\bea
|<\psi~ K_S~ |~ L ~ | B^0(p) > |^2~ &=&~e^{- \Gamma t} [1 - sin 2\beta ~
sin \Delta m t]~ |a (\eps^* \cdot p)|^2\\
|<\psi~ K_S~ |~ L ~ | \bar {B^0}(p) > |^2~ &=&~e^{- \Gamma t} [1
+ sin 2\beta ~
sin \Delta m t]~ |a (\eps^* \cdot p)|^2
\eea
which produces the asymmetry
\be
A_\beta~=~\frac {\Gamma( \bar {B^0} \to \psi~ K_S)~-~
\Gamma(  {B^0} \to \psi~ K_S)}{\Gamma( \bar {B^0} \to \psi~ K_S)
{}~+~\Gamma(  {B^0} \to \psi~ K_S)} ~=~ sin 2\beta ~
sin \Delta m t
\ee
where $\Delta m$ can be obtained by the
experimental value of $x_d$ (see (2.2)).
On the other hand, the integrated asymmetry is as follows:
\be
A_\beta ^{INT} ~=~\frac{x_d}{1+x_d^2} sin 2 \beta
\ee
with a dilution factor given by $\displaystyle \frac{x_d}{1+x_d^2}~=~0.46$.

{}From the the analysis of Section 3 on the constraints on the CKM parameters,
one obtains the lower bound \cite {lhc}
\be
sin~ 2 \beta >~0.16~,
\ee
for $m_t = 100 ~GeV$. The lower bound increases with $m_t$, for example
it gets the value $0.21$
for $m_t = 140~ GeV$ and $0.24$ for $m_t = 180~ GeV$.

Thus far we have not considered penguin diagrams. They would
contribute with an amplitude proportional to
$V^*_{tb} V_{ts} \simeq V_{ts} $.
 However the relative phase between the amplitude containing the mixing
and the W-internal emission diagram is
\be
arg~ \frac{V^*_{tb} V_{td}}{ V^*_{cb} V_{cs}} \simeq - arctn
\frac{\eta}{ 1-\rho} =
- \beta~,
\ee
which is identical to the phase difference between mixing and penguin
amplitudes:
\be
arg~ \frac{V^*_{tb} V_{td}}{ V^*_{tb} V_{ts}} \simeq =
- \beta~,
\ee
as one easily obtains from the CKM matrix in the Wolfenstein
parametrization. Therefore the presence of
penguin diagrams would not destroy the prediction (4.10);
on the contrary it would contribute constructively
to the asymmetry $A_\beta$.  We also observe that
the statistics for $A_\beta$ can be increased by adding more decay
channels with final states having the same $CP$ as
$ \psi~ K_S~$, such as
\be
B^0 \to \chi~ K_S~,~ \eta_c K_S~,
\ee
while final states with opposite $CP$ (e.g. $ \psi~ K_L$) can be also used to
increase the statistics (however they should be subtracted). In Table I
we list some of the processes that can be used to measure
$A_\beta$.

{\it Measurement of $sin 2\alpha$}

In order to measure $sin 2\alpha$ one can consider the decay
\be
B^0 \to \pi^+ \pi^-
\ee
whose branching ratio can be predicted, within the factorization approximation,
to be of the order of a few units $\times 10^{-4}$ \cite {wsb}. This result
is obtained by considering the external W emission
(spectator quark)  diagram, which is proportional to the product
$V^*_{ub} V_{ud}$.

For the amplitudes (4.16) we write:
\bea
<\pi^+ \pi^-| L | B^0 > &=&~+~A~e^{+i \gamma}\\
<\pi^+ \pi^-| L | \bar {B^0} > &=&~-~A~e^{-i \gamma}
\eea
where we have used our phase conventions. Computing these equations
at time $t$, similarly to the previous case we obtain the asymmetry:
\be
A_\alpha~=~\frac {\Gamma( \bar {B^0} \to \pi^+ \pi^-)~-~
\Gamma(  {B^0} \to \pi^+ \pi^-)}{\Gamma( \bar {B^0} \to \pi^+ \pi^-)
{}~+~\Gamma(  {B^0} \to \pi^+ \pi^-)} ~=~ sin 2\alpha ~
sin \Delta m t
\ee
As in the previous case, the statistics could be increased by
considering other related channels (see Table I).

Differently from $sin 2\beta$, $sin 2\alpha$ is not constrained by
the present bounds on the CKM matrix elements, as it can be seen
from Fig. 3: for example the value $sin 2\alpha~=~0$ is not excluded
by the data. The measurement of $sin 2\alpha$ is nevertheless very
important
theoretically. As a matter of fact, as shown in \cite {soares}
a measurement of $A_\beta$ alone would not allow to distinguish the
Standard Model (SM) from other theories. For example in
the superweak model \cite {sw}  (which is incidentally not yet ruled out since
the present experimental data on $\eps'/\eps: \eps'/\eps = (2.2 \pm 1.1)\times
10^{-3}$ \cite {pdb} are still compatible, within two standard deviations,
with the absence of direct $CP$ violations) one expects:
\be
A_\alpha~=~-~A_\beta~,
\ee
which is allowed, but not necessarily true in the Standard Model.

However the possibility to obtain clear indications on the value of
$sin~ 2\alpha$ from the $B^0 \to \pi^+ \pi^-$ decay channel depends on the role
of the penguin diagrams that have not been included in (4.17) and (4.18).
As a matter of fact, differently from the previous case, the penguin diagram
contributing to (4.16) has a different weak phase
($\dd e^{- i~\beta}$);
therefore, when added to (4.17) and (4.18),
it would change the prediction for
$A_\alpha$.

The  trouble with the penguin diagrams is that there is no
reliable way to compute them at the present; moreover, even
though they are small (for example there are indications that their role
in the decay width of $B^0 \to \pi^+ \pi^-$ is modest \cite {gronau1}) they
can nonetheless alter significantly the prediction for the asymmetry,
as shown, for example, in \cite {gronau2}.

{\it Measurement of $sin 2\gamma$}

$sin~2\gamma$ can be measured by $B_s$ ($=\bar {b} s)$ decays, e.g. by the
decay modes:
\be
B_s \to \rho^0~ K_S
\ee
In this case one finds:
\bea
<\rho^0(\eps, q) K_S(p')| L | B_s > &=&+A'~e^{+i \gamma}(\eps^* p)\\
<\rho^0(\eps, q) K_S(p')| L | \bar {B_s} > &=&+A'~e^{-i \gamma}(\eps^* p)
\eea
which gives the asymmetry:
\be
A_\gamma~=~\frac {\Gamma( \bar {B_s} \to  \rho^0~ K_S)~-~
\Gamma(  {B_s} \to  \rho^0~ K_S)}{\Gamma( \bar {B_s} \to  \rho^0~ K_S)
{}~+~\Gamma(  {B_s} \to  \rho^0~ K_S)} ~=~ sin 2\gamma ~
sin \Delta m t
\ee
As in the previous case, the statistics can be increased by
considering other channels (see Table I). However, the determination of
$sin 2 \gamma$ by this method presents the same
difficulties already mentioned for $A_\alpha$, i.e.  a non negligible role of
penguin diagrams. We shall discus an alternative method to
measure $sin 2 \gamma$ in the next Section.

{\it $c \bar {c}$ production from $B_s$}

The last class of processe we wish to consider in this Section includes
$B_s$ decays such as
\be
B_s \to \psi \phi,~ \eta_c \phi, \psi K_S
\ee
that at the quark level involve the production of $c \bar {c}$ pairs. They
produce an asymmetry:
\be
A_\delta~=~sin 2\delta ~ sin \Delta mt~,
\ee
where
\be
sin 2 \delta \sim - 2 \eta \lambda^2~.
\ee
Due to the value of $\lambda$ these asymmetries are probably too small to
 be observed in the next future.

\resection{Measuring $\gamma$ from charged $B$ decays}

Given the relevance of measuring, besides $sin 2\beta$,
also another element of the unitarity triangle,  we shall now examine
a possible way to
extract information from $CP$ violating asymmetries in charged $B$ decays
\cite {dunietz}.

Let us define
\be
D_{1,2}=\frac{D^0 \pm \bar {D^0}}{\sqrt {2}}
\ee
where $D_1$ ($D_2$) has $CP=-1$ ($CP=+1$). The amplitude for the decays
$B^+ \to  D_j K^+$ has the form
\be
A(B^+ \to D_{1,2} K^+)=\frac{1}{\sqrt 2}[A(B^+ \to D^0 K^+) \pm
A(B^+ \to \bar {D^0} K^+)]
\ee
while the amplitude for the decays
$B^- \to  D_j K^-$ is given by:
\be
A(B^- \to D_{1,2} K^-)=\frac{1}{\sqrt 2}[A(B^- \to D^0 K^-) \pm
A(B^- \to \bar {D^0} K^-)]
\ee

The amplitude $A(B^+ \to  D^0 K^+)$ is given by an internal  W-emission
diagram $\sim V^*_{ub}V_{cs} \sim e^{+i \gamma}$, while
the amplitude $A(B^+ \to \bar {D^0} K^+)$ is given by an external  W-emission
diagram $\sim V^*_{cb}V_{us}$, which is real. Therefore we can write
\bea
A(B^+ \to D_{1,2} K^+)=\frac{1}{\sqrt 2}[|A|e^{+i \gamma} e^{+i \lambda}
\pm |B| e^{+i \tau}]\\
A(B^- \to D_{1,2} K^-)=\frac{1}{\sqrt 2}[|A|e^{-i \gamma} e^{+i \lambda}
\pm |B| e^{+i \tau}]
\eea
where $\lambda$ and $\tau$ are strong phases ($\lambda \not= \tau$
because $\bar {D^0} K^+$ and $D^0 K^+$ have different isospin content).

The two relations (5.2) and (5.3) (for $D_j=D_1$)
can be represented geometrically
in the complex plane (Fig. 4) and define two triangles: ABC and ABD
respectively. We observe that, from (5.4) and (5.5),
AB= $A(B^+ \to \bar {D^0} K^+)~=~A(B^- \to  D^0 K^-)$ and
$\dd A(B^+ \to D^0 K^+)~=~e^{2 i \gamma}~A(B^- \to  \bar {D^0} K^-)$.
Therefore, by measuring $ \Gamma (B^+ \to D_1 K^+), \Gamma (B^- \to D_1 K^-),
\Gamma (B^+ \to \bar {D^0} K^+)=\Gamma (B^- \to D^0 K^-),
\Gamma (B^+ \to D^0 K^+) =
\Gamma (B^- \to  \bar {D^0} K^-)$
one can completely
reconstruct the
two triangles of Fig. 4 and measure the angle $\gamma$.

\resection{Conclusions}

While the measurement of the asymmetry in $B^0 \to \psi K_S$ represents by far
the cleanest way to observe $CP$ violations in $B$ decays, the mere
measurement of $sin 2 \beta$ would not allow a clear distinction between
Standard Model and the possible effects of new physics. A simultaneous
measurement of two angles of the unitarity triangle is necessary, but,
probably, one has to consider not only neutral $B$ decays into CP
eigenstates, but charged B decays as well.

In any event, hadron colliders and,
in particular, a dedicated UNK B-factory would represent an excellent
tool to investigate this fascinating area of the elementary particle physics
given the copious production of beauty particles that is expected
and the possibility they offer to investigate different and complementary
decay channels.

\newpage
\begin{center}
  \begin{Large}
  \begin{bf}
  Table Caption
  \end{bf}
  \end{Large}
\end{center}
  \vspace{5mm}
\begin{description}
\item [Table I]
Quark and hadron processes contributing to the $CP$ violating asymmetry
$\displaystyle
A_\Lambda = \frac {\Gamma( \bar {B^0} \to f) - \Gamma ({B^0} \to f)}
{\Gamma (\bar {B^0} \to f) + \Gamma ({B^0} \to f)}= - sin 2 \Lambda sin
\Delta m t$. Penguin diagrams may contribute constructively to
the asymmetry $A_\Lambda$ or destroy the prediction.
\end{description}

\newpage
\begin{table}
\begin{center}
\begin{tabular}{l c c c c }
& & {\bf Table I} & & \\ & & & & \\
 \hline \hline
Quark Process & Hadron Process & $\Lambda$ & CP(f) & Penguin diagrams role \\
\hline & & & & \\
${\bar b} \to c {\bar c} {\bar s}$ & $B_d \to \psi K_S$
& $- \beta$ & -1 & Constructive contribution \\
 & $B_d \to  \chi K_S  $ & $- \beta$ & -1 & "~~~~" \\
 & $B_d \to  \eta_c K_S  $ & $- \beta$ & -1 & "~~~~" \\
 & $B_d \to  \psi K_L  $ & $+ \beta$ & +1 & "~~~~" \\
 & $B_d \to  \chi K_L  $ & $+ \beta$ & +1 & "~~~~" \\
 & $B_d \to  \eta_c K_L  $ & $+ \beta$ & +1 & "~~~~" \\
 & & & & \\
${\bar b} \to c {\bar c}{\bar d}$ & $B_d \to D^+ D^-$ & $+ \beta$ & +1 &
"~~~~" \\
 & $B_d \to D^0 \bar {D^0}$ & $+ \beta$ & +1 & "~~~~" \\
 & & & & \\
${\bar b} \to {\bar s}$(penguin)& $B_d \to K_S \omega$ &$-\beta$ & -1&"~~~~" \\
 & $B_d \to K_S \rho^0$ & $-\beta$ & -1&"~~~~" \\
 & $B_d \to K_S \Phi$ & $-\beta$ & -1&"~~~~" \\
 & $B_d \to K_S \omega$ & $+\beta$ & +1&"~~~~" \\
 & $B_d \to K_S \rho^0$ & $+\beta$ & +1&"~~~~" \\
 & $B_d \to K_S \Phi $ & $+\beta$ & +1&"~~~~" \\  & & & & \\ \hline
 & & & & \\
${\bar b} \to u {\bar u} {\bar d}$ & $B_d \to \pi^+ \pi^-$ &$-\alpha $ & +1&
Possible \\
& $B_d \to \pi^0 \pi^0$ &$-\alpha $ & +1&
" \\
& $B_d \to \omega \pi^0$ &$-\alpha $ & +1&
" \\
& $B_d \to \rho \pi^0$ &$-\alpha $ & +1&
" \\
& $B_d \to p {\bar p}$ (s-wave) &$ +\alpha $ & -1&
" \\
& $B_d \to p {\bar p}$ (p-wave) &$ -\alpha $ & +1&
" \\ & & & & \\ \hline & & & & \\
${\bar b} \to u {\bar u} {\bar d}$ & $B_s \to \rho K_S $ & $-\gamma $ & -1&
 Possible  \\
& $B_s \to \omega K_S $ & $-\gamma $ & -1&
" \\
& $B_s \to \rho K_L $ & $+\gamma $ & +1&
" \\
& $B_s \to \omega K_L $ & $+\gamma $ & +1&
" \\ & & & & \\
\hline \hline

\end{tabular}
\end{center}
\end{table}

\newpage
\begin{center}
  \begin{Large}
  \begin{bf}
  Figure Captions
  \end{bf}
  \end{Large}
\end{center}
  \vspace{5mm}
\begin{description}
\item [Fig. 1]
The box diagram; solid lines represent quarks.
\item [Fig. 2]
The unitarity triangle.
\item [Fig. 3]
Allowed region in the $(\rho, \eta)$ plane from $\eps_K$ measurement
(solid lines), from $|V_{ub}/V_{cb}|$ (dotted lines) and
from $B^0 - \bar {B^0}$ mixing (dashed lines).
Also reported is the unitarity triangle of
Fig. 2.
\item [Fig. 4]
Measurement of $2\gamma$ by charged $B$ decays.
The different complex amplitudes are represented  by vectors
comprising the two triangles ABC and ABD as follows:
AB= $A(B^+ \to \bar {D^0} K^+)$~=~$A(B^- \to  D^0 K^-)$;
BD= $A(B^- \to \bar {D^0} K^-)$;
AD= ${\sqrt 2} A(B^- \to  D_1 K^-)$;
AC= ${\sqrt 2} A(B^+ \to D_1 K^+)$;
BC= $A(B^+ \to D^0 K^+)$.
\end{description}

\end{document}